# Accurate Analysis of the Edge Taper Influence on the Performance of Hemielliptic Lens Antennas

Artem V. Boriskin [#1], Ronan Sauleau [*2], Alexander I. Nosich [#3]

[#] *Institute of Radiophysics and Electronics NASU, vul. Proskury 12, Kharkiv 61085, Ukraine*

[1] a_boriskin@yahoo.com

[*] *Institut d'Electronique et de Télécommunications de Rennes, Université de Rennes 1, UMR CNRS 6164, 35042 Rennes cedex, France*

*Abstract* — **Correlation between the lens extension size and the broadside directivity of a hemielliptic dielectric lens antenna (DLA) fed by a primary feed with variable radiation pattern is studied in accurate manner. The problem is considered in two-dimensional formulation and solved numerically using in-house software based on the Muller boundary integral equations (MBIE). Our results highlight the key role of the edge taper which can be defined for DLAs similarly to the theory of reflector antennas. A new feature revealed is the relation between the optimal edge taper needed to achieve the highest possible directivity and the permittivity of the lens material.**

## I. Introduction

Both parabolic reflectors and elliptic lenses are designed to collect the parallel rays into a focus [1]. By reciprocity, to use this focusing ability in the full manner in the emitting mode one needs a feed capable of providing a uniform illumination of the reflector or the lens front profile. For realistic feeds, the spillover and illumination losses are inherently present [2]: the former is associated with the power that misses reflector/lens whereas the latter is due to a non-uniform illumination of the reflector/lens front part (Fig. 1). The optimal antenna performance can be achieved if a proper edge taper (i.e. reduced illumination) is provided. For reflector antennas the recommended ratio of power at the reflector edge and at the centre is around -10 dB. As far as we know, verification of a similar recommendation for elliptical DLAs has never been published. If properly introduced, the edge taper analysis can help to answer the question about the optimal lens extension size discussed in [3-5]. Note that trustable results in such a study can only be obtained with application of an adequate simulation tool capable for accurate description of the resonance phenomena intrinsic to compact size lenses [6]. This is because both the electrical size and the focal distance of elliptic DLA are usually much smaller than that of reflectors [3], and thus the feed is never far away from the lens. Furthermore, unlike parabolic reflectors, any dielectric lens is, in fact, an open dielectric resonator that is capable of supporting resonant modes. The quality factors of such modes depend on the lens parameters (shape, size, and permittivity) and can achieve rather high values for lenses made of dense materials such as silicon. If excited, internal resonances strongly affect the performance of DLAs [6, 7]. Finally, for DLAs, the focal distance and thus the favourable feed location depend on the lens material. This happens because, in geometrical optics approximation, the eccentricity of elliptic lens is determined by its material permittivity [1]. These strong differences between reflector antennas and DLAs make the -10 dB optimal edge taper a questionable recommendation and call for additional study aimed at clarification of the role of edge illumination.

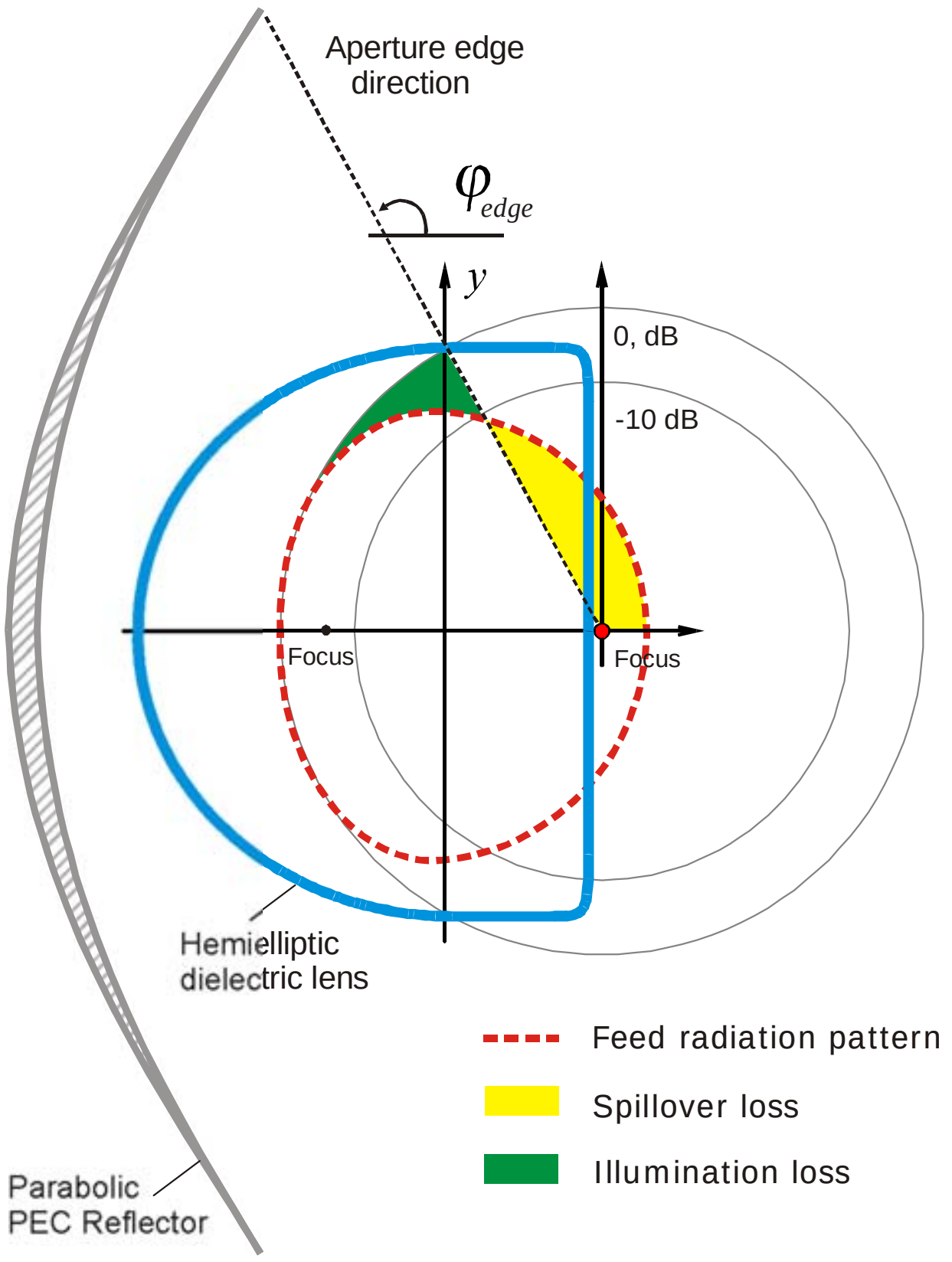

Fig. 1. Schematic diagram illustrating losses associated with a non-uniform illumination of the aperture of a parabolic reflector and a hemielliptic lens.

## II. Mathematical Model

In 2D, a DLA can be modelled as a homogeneous dielectric cylinder fed by a primary feed with a given radiation pattern. For DLAs, the feed is usually fixed directly to the lens flat bottom. To account for this, it is assumed radiating in a media with the same permittivity as the lens, $\varepsilon$. The scattering problem is solved numerically using the MBIE-based algorithm capable of accounting for all wave effects (multiple internal reflections, total reflections, surface waves, etc.) in a full manner. Details of the mathematical approach can be found in [6] whereas the description of the DLA model, that is important for interpretation of the results, is provided below.

In the paper, the lens profile is a combination of hemiellipse (front part) with eccentricity chosen in accordance with the GO focusing rule, i.e. $e = \varepsilon^{-1/2}$, and hemisuperellipse (rear part) smoothly joined at the points $(0, \pm a)$, where $a$ is the minor hemiaxis of the ellipse (Fig. 2). Note that the lens focusing ability is determined only by its elliptical front part, therefore these junction points coincide with the edge of the lens aperture (Fig. 1).

The feed is simulated by a complex source point (CSP) that is a current line located in complex space [8]. CSP is an attractive model of an aperture feed because its field is a unidirectional beam whose waist is controlled by the value of the imaginary part of the CSP coordinate [7]; it behaves like a Gaussian beam in the paraxial zone, whereas in the far zone CSP field smoothly transforms into a cylindrical wave and thus (in contrast to a Gaussian beam) satisfies the Sommerfeld radiation condition at infinity. The notations and the near-field map of the CSP are given in Fig. 3, and its far-field asymptotic is given by:

$$U^{in}(r,\varphi) \sim (2/i\pi kr)^{-1/2} \cdot \exp(ikr) \cdot \exp[kb\cos(\varphi-\beta)], \qquad (1)$$

where $\varphi$ is the polar angle of the observation point.

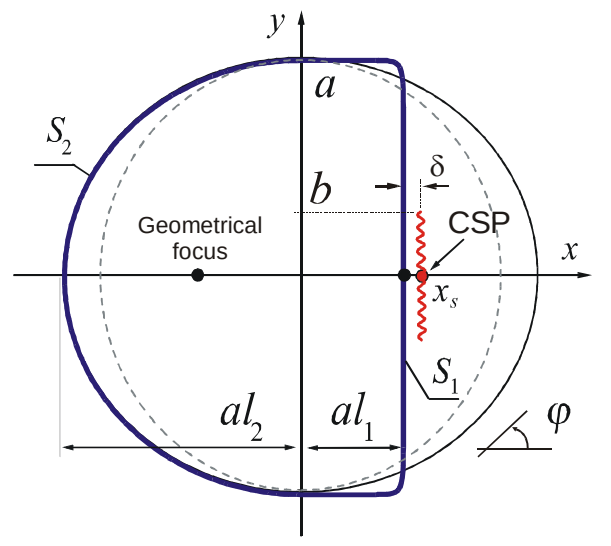

Fig. 2. Geometry and notations of a 2D model of a hemielliptic DLA fed by a CSP located close to the lens flat bottom ($\delta = \lambda_0/20$).

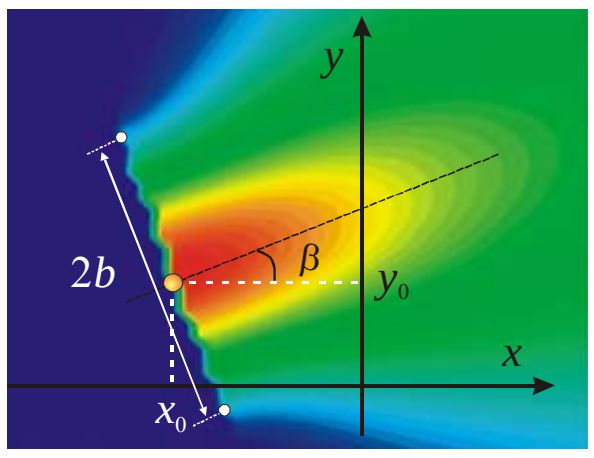

Fig. 3. Notations and near-field pattern of a CSP radiating in free space ($kb$ = 4.0). White dots denote the branching points in the real space and the black dashed line shows the orientation of the radiated beam axis.

For reflector antennas having a large electrical size, the edge illumination is typically defined as the ratio in the source power radiated in the edge direction and in the broadside [2].

$$A = 20\log\left(U(\varphi_{edge})/U(\varphi_{bdside})\right)\big|_{r\to\infty} \qquad (1)$$

where $U$ is the field amplitude (in 2-D, this is $E_z$ and $H_z$ for the $E$- and $H$-polarizations, respectively) and $\varphi_{bdside}$ is the broadside (forward) direction. Such a definition, based on the far-field radiation pattern of the feed, is convenient because it clearly explains the physical origins of the losses and simplifies engineering specifications for feeds. For hemielliptic lens fed by the CSP located and oriented as shown in Fig. 2, a closed-form expression for the edge illumination is given by:

$$A \approx -8.68\ kb\sqrt{\varepsilon}\,(1+\cos\varphi_{edge}) \qquad \text{[in dB]}. \qquad (2)$$

If the lens is cut through its rear focus (hemielliptic DLA), then the normalized lens extension and the "edge direction" are defined as $l_1 = (\varepsilon-1)^{-1/2}$ and $\cos\varphi_{edge} = \varepsilon^{-1/2}$, respectively.

For DLAs whose size is often only a few wavelengths the far-field definition of edge illumination is less applicable and should be replaced by the one based on the near-fields: e.g. defined as the ratio of the incident field intensity at the "edge" of the lens aperture and in its center:

$$A = 20\log[U(0,a)/U(0,0)] \qquad \text{[in dB]}. \qquad (3)$$

The difference between these two definitions is well seen in figures given in Section III (Figs. 4-6), where two curves for the edge illumination defined via the far and near fields are indicated. Note that the same edge illumination can be provided by feeds with different radiation patterns and therefore the numerical results presented in the paper should be considered as reference ones and additional correction may need to be applied for other feeds depending on their radiation patterns.

The radiation characteristic considered as a measure of collimation ability of the lens is the broadside directivity defined as $D = 2\pi|U(\varphi_m)|^2/P_{tot}$, where $P_{tot} = \int_0^{2\pi}|U(\varphi)|^2 d\varphi$ is proportional to the radiated power integrated over all directions. Note that the directivity of the CSP radiating into infinite medium is $D_e = \exp(2\sqrt{\varepsilon}kb)\,/\,I_0(2\sqrt{\varepsilon}kb)$, where $I_0$ is the modified Bessel function. This function is represented by the curve marked with black circles in Figs. 4-6.

## III. Numerical Results

Our analysis shows that adjustment of edge illumination enables one to improve the DLA broadside (forward) directivity roughly by factor 2 (Fig. 4-6). As it is seen, for lenses of both sizes ($a = 2\lambda_0$ and $a = 4\lambda_0$) and made of different materials ($\varepsilon = 2.53 \div 11.7$) the directivity grows proportionally to the CSP aperture (controlled by parameter $kb$) until some optimal edge taper is achieved and then almost monotonically goes down. The maxima of directivity are indicated by vertical dotted lines. The optimal edge taper can be determined by the intersection of these vertical lines with the inclined dashed lines associated with the right axis representing the edge illumination level.

As one can see, the optimal value of -10 dB, recommended for reflector antennas where the far-field edge-illumination definition is common, can still be applied for DLAs if the near-field definition is used. More precisely, this recommendation is uniformly applicable to lenses made in Rexolite, as well as to denser materials like quartz and silicon in *E*-case. For *H*-case it must be modified in favor of -12 dB and -20 dB values for the quartz and silicon lenses, respectively. However, the maximum of the *D*(*kb*) curve is broad so the tolerance in the optimal edge taper value is quite large and can be estimated within the ± 5 dB range.

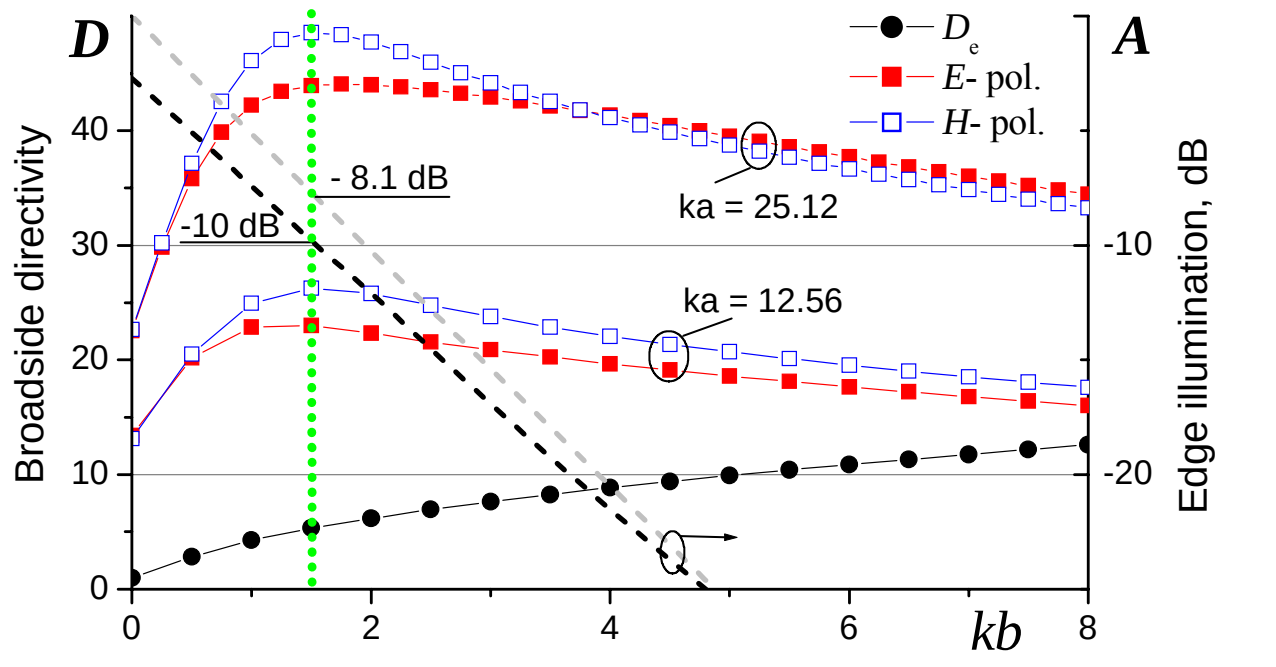

Fig. 4. Broadside directivity (left scale) and edge illumination (right scale) of the cut-through-focus rexolite hemielliptic DLA ($\varepsilon = 2.53$, $l_1 = 0.8$, $l_2 = 1.286$) vs. CSP aperture width. The inclined dashed lines associated with right axis indicate the edge illumination defined via near fields (black) and far fields (grey). The vertical dotted line indicates the maximum value of the directivity and is plotted to help estimating the optimal value of edge taper.

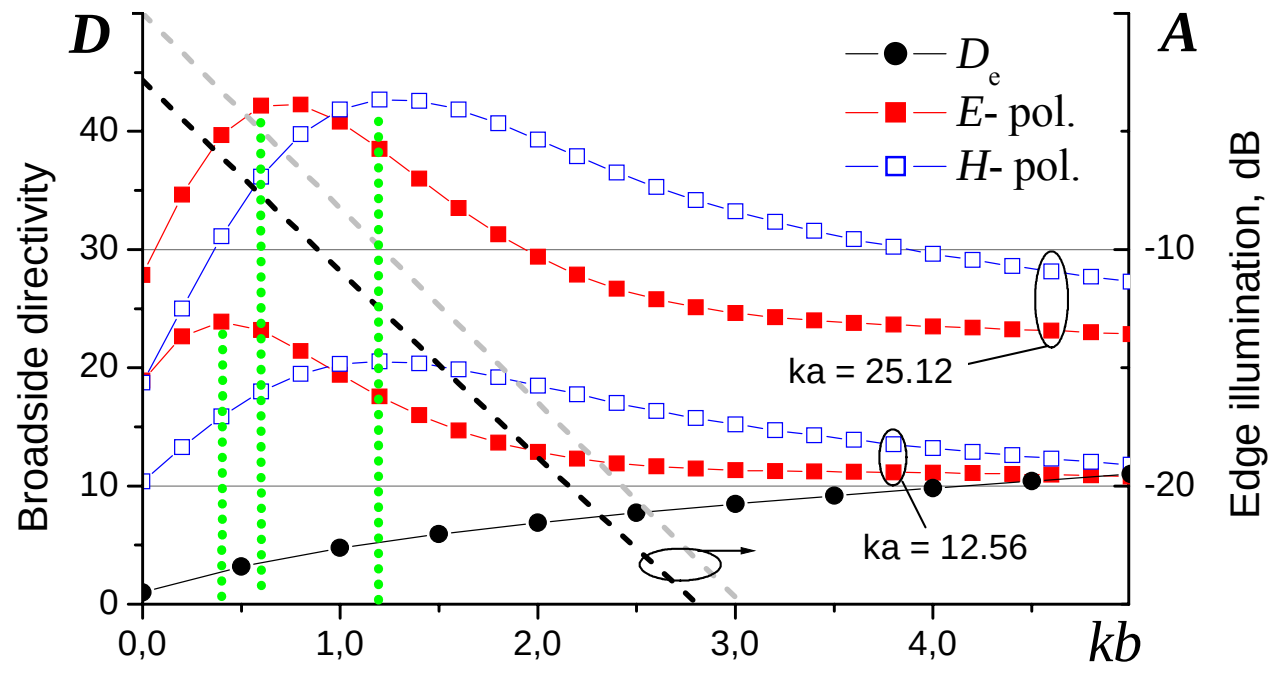

Fig. 5. The same as in Fig. 4 for the quartz DLA: $\varepsilon = 3.8$, $l_1 = 0.6$, $l_2 = 1.165$.

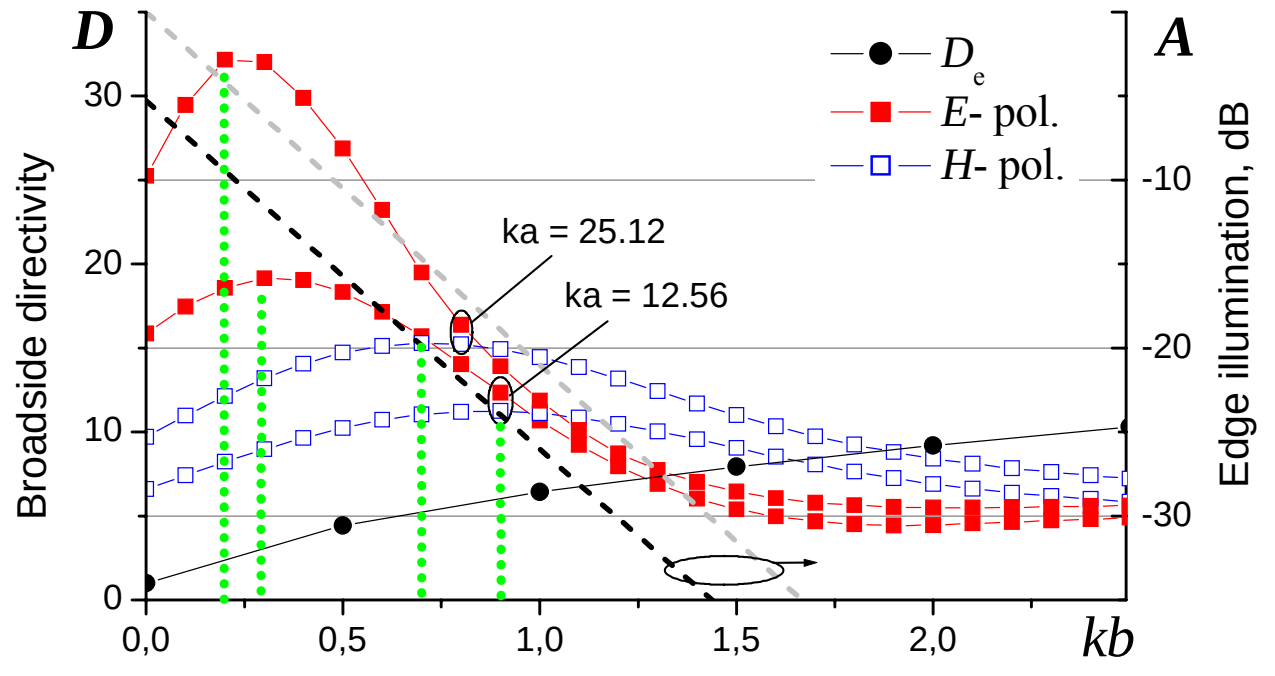

Fig. 6. The same as in Fig. 4 for the silicon DLA: $\varepsilon = 11.7$, $l_1 = 0.3$, $l_2 = 1.046$.

Additional information about the collimation properties of the DLAs can be extracted from the relief maps of the broadside directivity computed for lenses with variable extension size fed by CSPs with variable apertures (Fig. 7-9). As the edge taper depends on both parameters, the right axis scale corresponds only to the values of $l_1$ marked by the vertical dashed lines. The non-monotonic behavior of directivity highlights important role of internal reflections in the electromagnetic behavior of compact-size dielectric lenses. For instance, for silicon lens strong internal resonances are become apparent in the form of deep periodic valleys running along the vertical axis for a number of the lens extension values (Fig. 9).

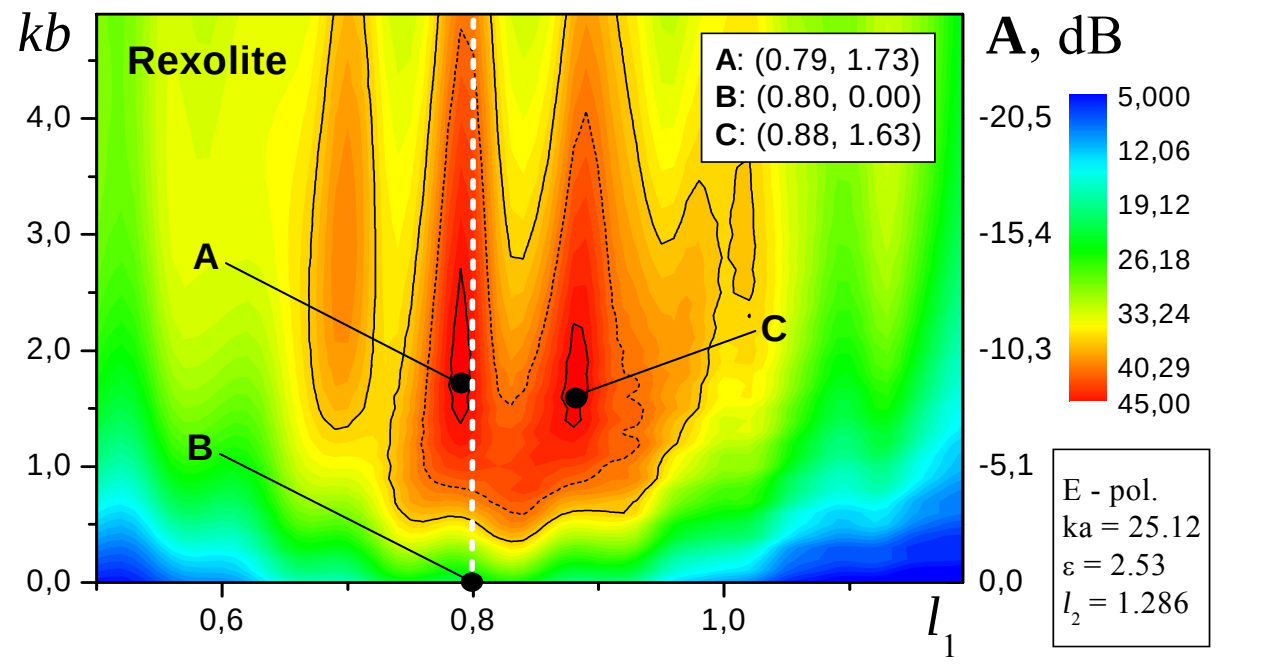

Fig. 7. Broadside directivity vs. normalized lens extension ($l_1$) and CSP aperture width (*kb*). For clarity only top 2%, 10% (dotted line), and 20% grids are shown. The right axis gives the value of edge taper for the lens cut through the focus (the corresponding value of the lens extension is indicated by the vertical dashed line). The marks correspond to the far-field radiation patterns given in Fig. 10.

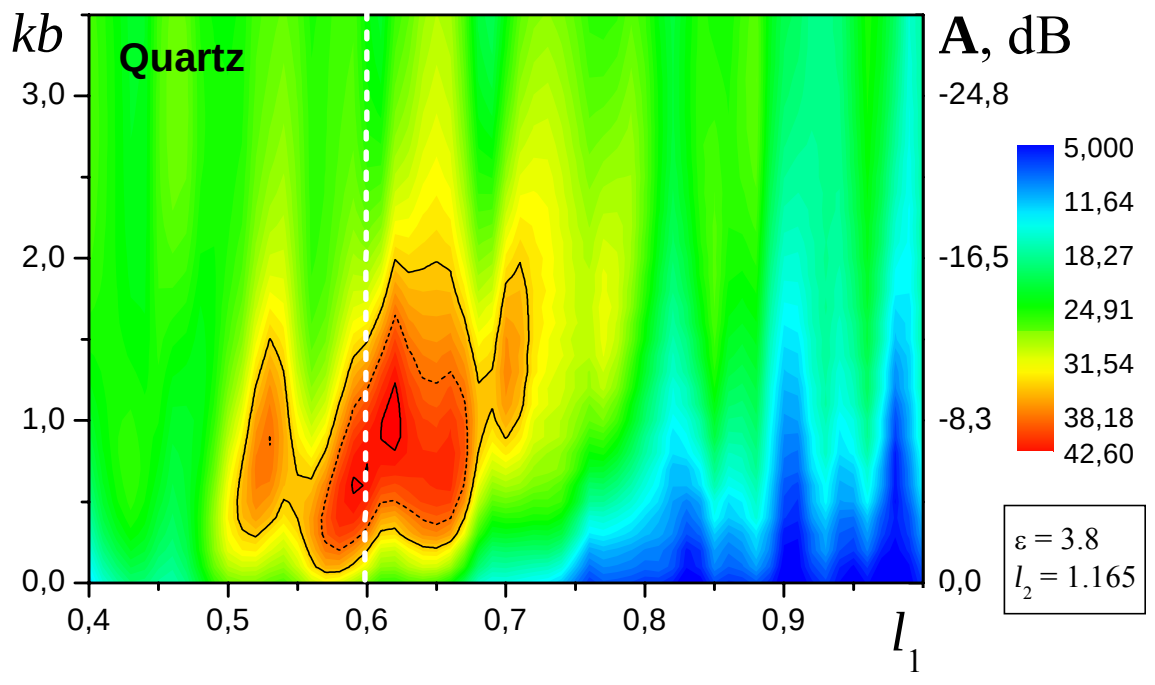

Fig. 8. The same as in Fig. 7 for the quartz lens: $l_1 = 0.6$, $l_2 = 1.165$.

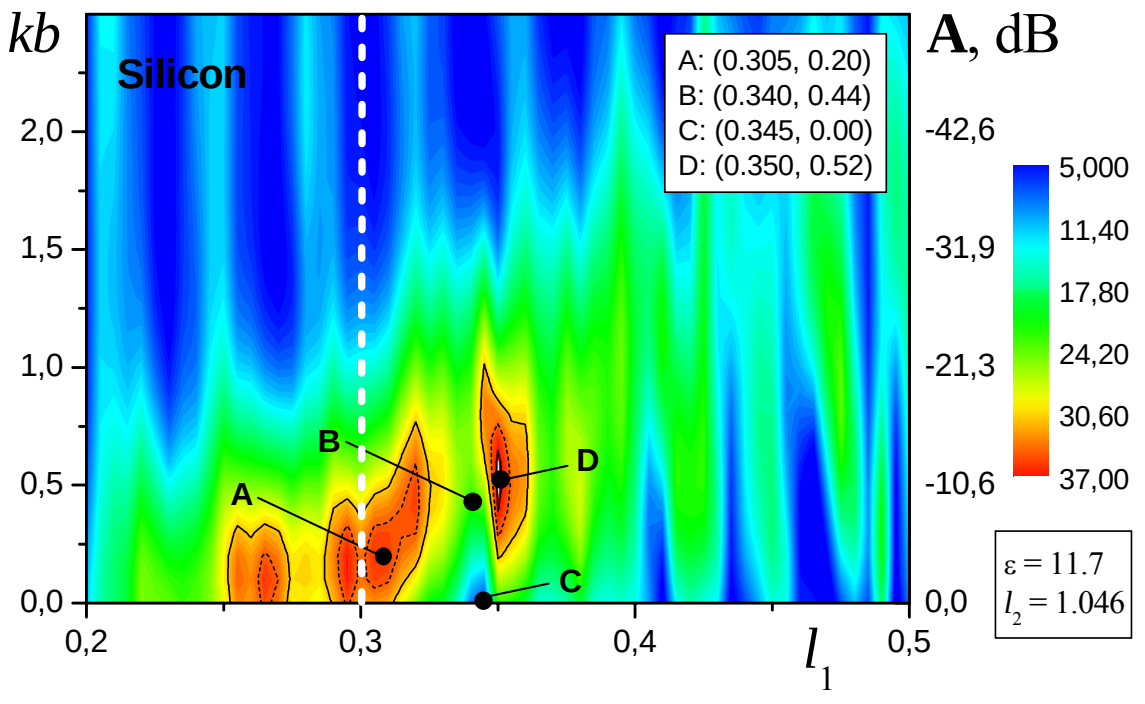

Fig. 9. The same as in Fig. 7 for the silicon DLA: $l_1 = 0.3$, $l_2 = 1.046$. The marks correspond to the far-field radiation patterns given in Fig. 11.

Note that the valleys running in the top (for the larger values of *kb*) are associated with Fabry-Perot effect or bouncing of the internal field that is proven by their periodicity, whereas the aperiodic ones in the figure bottom (for smaller *kb*) are associated with the so-called half-bowtie (HBT) resonances studied in [6]. The important observation from Fig. 7-9 is that a "good choice" of the lens extension itself does not guarantee the highest directivity. This value can be considerably improved by tuning the feed radiation and hence obtaining the optimal level of the lens aperture edge illumination. However, these recommendations are relevant only if the frequency is far from an internal resonance that can significantly affect the performance of the DLAs [6].

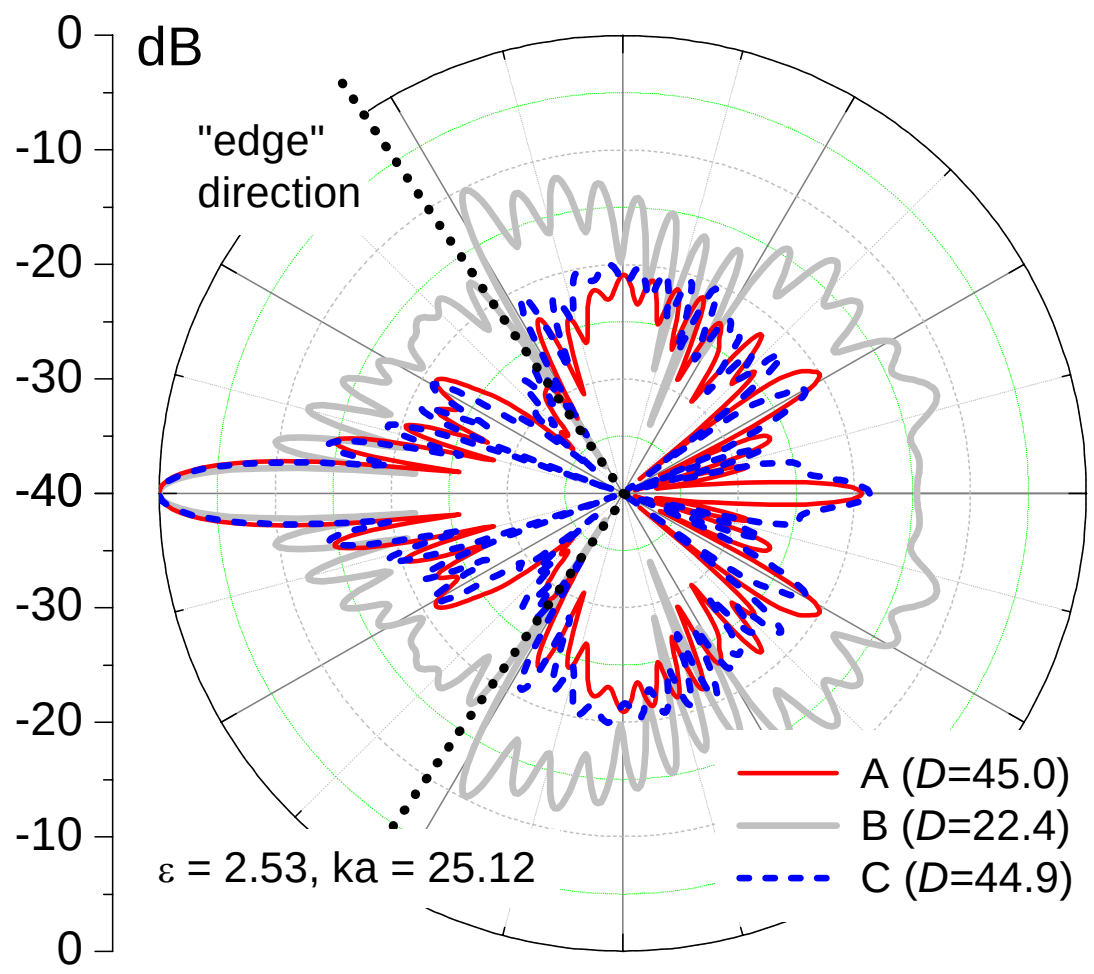

Fig. 10. Normalized far-field radiation patterns for the rexolite DLA excited by *E*-polarized CSPs whose parameters correspond to the relevant marks in Fig. 7. The values of the broadside directivity for each of geometries are given in the legend. *Note the significant difference in the side-lobe level and spillover radiation observed for the cut-through-focus lens excited by the omnidirectional feed and DLAs with optimal edge taper.*

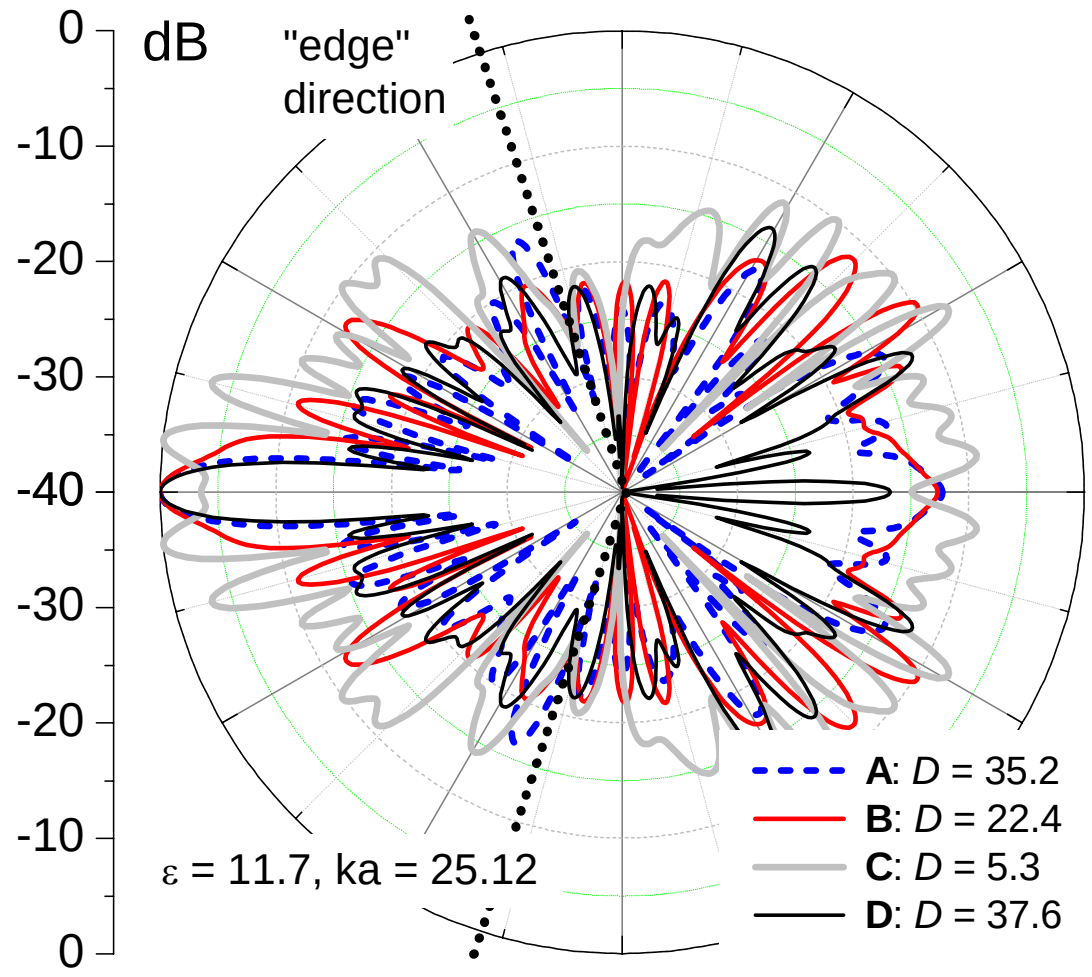

Fig. 11. The same as in Fig. 10 for the silicon DLA. For parameters see the relevant marks in Fig. 9. *Note the splitting of the main beam for C-curve that appears due to the HBT resonance excitation and the presence of inherent resonant side-lobes for B-curve.*

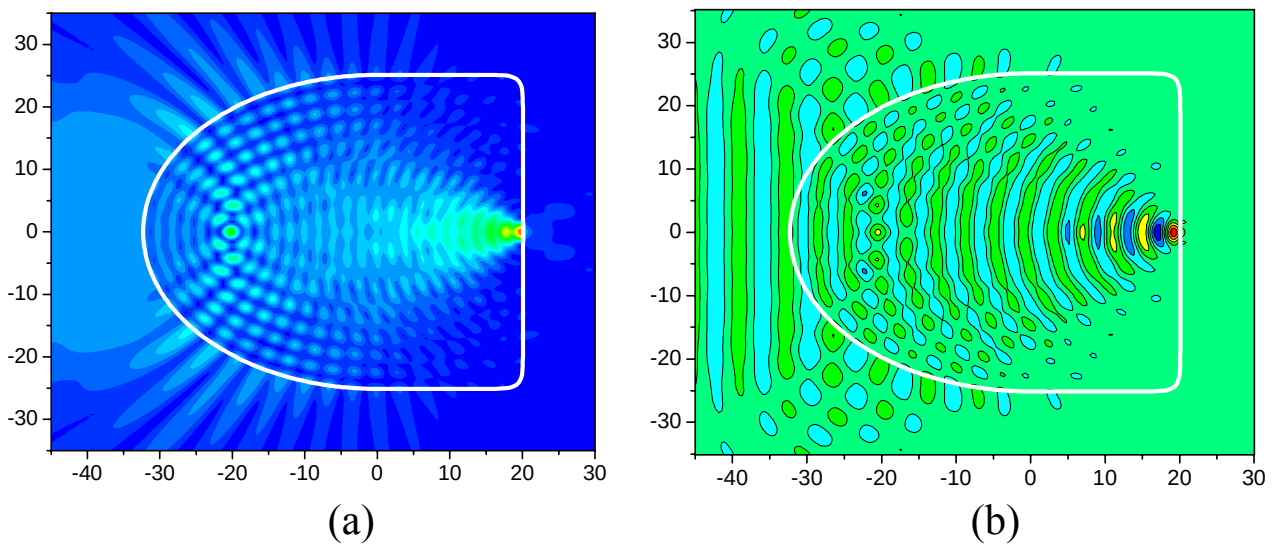

Fig. 12. Near-field amplitude (a) and phase (b) patterns of the rexolite DLA excited by the *E*-polarized CSP source. The lens and feed parameters correspond to Fig. 7, mark A).

The far-field radiation patterns presented in Figs. 10 and 11 are plotted for DLAs whose parameters correspond to characteristic points marked in Figs. 7 and 9, respectively. Here, Fig. 10 visualizes the formation of the main beam via suppression of the side-lobes (minimization of the spillover losses), whereas Fig. 11 also highlights the effect of the main beam degradation due to excitation of an HBT resonance. Finally, the near-field intensity and phase distributions for the rexolite DLA with the optimal edge illumination (Fig. 7, mark A) are shown in Fig. 12 to demonstrate the formation of a locally-plane wave with a uniform phase distribution in the output aperture of the lens.

## IV. Conclusions

The performance of elliptical DLAs with variable back-side extension and primary feed pattern has been studied in order to determine the optimal range of edge taper illumination needed to achieve the highest possible directivity. It has been demonstrated that the far-field definition of edge taper typically used for reflector antennas gives the optimal value of -7–8 dB instead of -10 dB often referred as the optimal one. However, the latter becomes true if the near-field definition is applied. Moreover, it was found that, unlike reflector antennas, the optimal edge taper for DLAs depends on the lens material and polarization of the primary feed. Finally, it was demonstrated that optimal edge taper does not prevent from excitation of internal resonances that can be excited even if a directive feed is used.

More details on the discussed problem are available in [9].

## Acknowledgment

This work was supported in part by the joint projects between IRE NAS Ukraine, on the one side, and the CNRS, MENESR, and MAEE France, on the other side. The first author was also supported by the Foundation Michel Métivier and by NATO via Grant NIKR.RIG.983313.